# Automating Sexual Injustice:

## Epistemic Injustice in Fembot Design and Feminist Directions for Equitable HRI

Surabhi Bhardwaj
The Leverhulme Centre for
the Future of Intelligence
(LCFI)
University of Cambridge
Cambridge, United Kingdom
sb2629@cam.ac.uk

## ABSTRACT

Current AI-enabled female sex robots, or "fembots," are primarily designed to simulate female sexual responses through a lens of male-centric bias and pornographic stereotypes. This paper analyses fembot development as a failure in equitable robotics, arguing that these machines perpetuate "epistemic injustice" by prioritizing male hedonistic fantasies over empirical truths of female sexual experience in their design decisions. Drawing on Miranda Fricker's framework of testimonial and hermeneutical injustice, this analysis demonstrates how fembot interfaces discredit women's lived sexual knowledge and empirical research on female sexual physiology while privileging male-centred fantasies. This paper proposes three Feminist Design Directions — empirical grounding, epistemic plurality, and active consent modelling — grounded in Donna Haraway's concept of "Situated Knowledge" and accompanied by concrete evaluation criteria. These directions aim to facilitate a transition toward evidence-based intimate AI that prioritizes epistemic justice, mutuality, and inclusive design for marginalized users including disabled, neurodivergent, and LGBTQ+ communities.

## 1   INTRODUCTION: ROBOTS FOR SEXUAL WELL BEING

As AI technologies permeate the intimate domains of human experience, the design of "sexbots", specifically fembots, emerges as a critical site of socio-technological inquiry. Sexbots are life-sized robotic representations of the human body, equipped with AI and silicone skin to mimic the sensation of human touch, and are used for sexual purposes such as sexual simulation and gratification [1, 2]. Fembots are a subcategory of sex robots designed to simulate the female body and sexual responses: they are programmed to mimic female reactions, for example, moaning when touched, simulating orgasms, and giving verbal affirmations of pleasure, thereby producing a mechanical simulation of female desire [3, 4]. Companies such as RealDollX and TrueCompanion allow users to customize nearly every aspect of a fembot's physical appearance and personality, including body shape, size and personality traits [4]. While a growing body of scholarship in equitable design and human–robot interaction (HRI) prioritises design goals that serve marginalised communities and promote dignity, the current development of fembots is reduced to a narrow framework of heterosexual male hedonism. Fembots represent a rapidly growing segment of the broader sextech industry, which was valued at $43 billion globally in 2024 [5], underscoring the urgency of subjecting their design assumptions to rigorous scrutiny. This paper identifies fembot development as a domain where gender bias and harmful stereotypes are technically automated, resulting in significant negative implications for sexual wellbeing and gender equity.

This paper argues that current fembots represent a profound failure in "Equitable HRI" by reinforcing epistemic injustice, a concept developed by philosopher Miranda Fricker [6] to describe the harm done to individuals specifically in their capacity as knowers. By misrepresenting women's sexual physiology, fembots encode reductive, male-centred scripts that undermine women's credibility as sexual knowers. Although marketed by techno-optimists as tools for therapeutic sexual "education" [7], the underlying interface design often prioritizes pornographic simulation over epistemic accuracy, which ultimately exacerbates sexual ignorance. The paper then proposes Feminist Design Directions and examines their real-world limitations and constraints.

## 2. LITERATURE REVIEW: EXISTING DEBATES ON FEMBOTS

Techno-optimistic scholars claim that sex robots deliver two key benefits: intense sexual pleasure and educational skill-building in sexual technique and intimacy, with proponents arguing that regular engagement can enhance relationship satisfaction [10, 11, 7]. However, this paper argues that these two goals are fundamentally misaligned in current fembot design. When functionality is tailored primarily to heterosexual male users, design prioritises "spectacle" over "truth" [12], fembots are frequently trained on pornographic scripts rather than empirical sexual research [13]. Williams [12] describes the pornographic orgasm as a "frenzy of the visible," a spectacle designed to confirm male sexual success, a logic that fembots inherit and automate. In practice, this produces interfaces in which users can trigger orgasms on command regardless of relational interaction [14], meaning that fembots do not educate; they perpetuate a male-centric model of sex that ignores the lived reality of female embodiment.

Feminist critics raise substantial challenges to the techno-optimist framing on ontological, ethical, and virtue-ethical grounds. Richardson [15] argues that fembots reinforce a patriarchal ontology in which femininity is equated with hypersexual servitude and constant availability — these machines are simulations of femininity designed to gratify, obey, and never resist, exacerbating the objectification of women and generating social expectations that real women should mirror robotic submissiveness [16, 15]. Gutiu [9] introduces the concept of the "roboticization of consent," whereby a robot's refusal can be overridden by voice command, rendering consent meaningless and undermining its moral force. Building on this, Kaufman [14] warns that representing women as perpetually consenting may reinforce harmful expectations in real-life sexual dynamics, potentially exacerbating rape culture. From a virtue ethics perspective, repeated engagement with submissive machines risks eroding users' moral sensibilities, fostering callousness, entitlement, and a decreased capacity for empathy [3, 17]; Oleksy and Wnuk [16] further caution that degrading behaviours normalised through sexbot interaction may spill over into human relationships, making fembots catalysts for habituation rather than harm reduction.

While these ontological, ethical, and virtue-based critiques establish a robust critical foundation, a key dimension remains underexplored in the existing literature: the feminist epistemic analysis of the sexual knowledge fembots produce and disseminate, and the specific harms of testimonial and hermeneutical injustice that follow.

## 3. FEMBOTS AS INSTRUMENTS OF EPISTEMIC INJUSTICE

This epistemic dimension manifests concretely in the interface design of fembots themselves (Figure 1).

## 3.1 Testimonial Injustice: Marginalizing Embodied Knowledge

In HRI design, testimonial injustice occurs when the lived experience of a marginalized user group is excluded from the system's behavioural engine or speech programs [15]. Current fembots enact this by elevating pornographic fantasy over empirical sexual science. Their sexual scripts are derived from pornographic stereotypes that prioritise the performance of male sexual success — exaggerated female moaning, submissive vocalisation, and visible climax — over grounding in the lived reality of female sexual experience [12].

While landmark research [18, 13, 11] demonstrates that non-penetrative stimulation is the primary physiological pathway to orgasm for the vast majority of women — with penile-vaginal intercourse (PVI) accounting for orgasm in less than 2% of women [13] — fembots are almost exclusively programmed to simulate orgasm via penetration. By automating "orgasm-on-penetration" and "orgasm-on-command" [14], developers treat women's authentic, embodied accounts of sexual pleasure as epistemically irrelevant. This creates a user-experience (UX) bias that validates male-authored fantasies while effectively ignoring the physiological reality of women's sexual lives [19, 20]. When women's status as "knowers" of their own bodies is discredited in the design phase, the resulting technology serves a hedonistic utility for one group while enacting an epistemic erasure of another.

## 3.2 Hermeneutical Injustice: Narrowing Sexual Scripts

Hermeneutical harm occurs when a technology limits the collective interpretive resources available to understand a human experience. Fembots compound this by automating the "Orgasm Gap" — a cultural phenomenon where male pleasure is treated as the normative standard of sexual success while female pleasure is rendered optional or secondary [21, 11, 22].

Additionally, hermeneutical injustice extends beyond sexual pleasure to the erasure of consent as a legible concept. Current fembot design reduces consent to a mechanical setting rather than a communicative act, structurally encoding compliance as the default condition of intimate interaction. This operationalises what Gutiu [9] terms the 'roboticization of consent' — a design logic in which refusal is a temporary state to be bypassed rather than a boundary to be respected [14]. Srinivasan [23] suggests that our sexual imaginations are shaped by the political and technological structures we inhabit. If the main "sexual script" available to a user is one where a partner never tires, never says "no,"[9] and always performs an exaggerated climax [24], the user loses the interpretive tools required to navigate the messy, non-linear reality of human consent. This design choice miseducates the user, providing a false sense of sexual expertise grounded in simulation rather than mutuality [7]. Habitual engagement with these submissive scripts may also erode the user's moral sensibilities, fostering character traits such as entitlement and a decreased capacity for empathy — vices that virtue ethicists warn may spill over into human relationships [17].

## 3.3 Intersectional Erasure as Epistemic Harm

Additionally, the epistemic harm of fembots is profoundly intersectional. As Benjamin [25] notes, AI often functions as a "New Jim Code," automating social hierarchies. Current fembot design reinforces a hierarchy of "knowable" bodies that is overwhelmingly white, thin, heterosexual and able-bodied [23]. This renders the lived experiences of Black, Brown, disabled, and LGBTQ+ bodies epistemically invisible in the realm of intimate HRI. When the only available body configurations are white and able-bodied, and the only legible sexual script is heterosexual and penetration-centric, the technology not only fails to include marginalised users, it also actively encodes their exclusion as the default condition of intimate HRI. Disabled users whose sexuality may require non-standard interaction modalities, neurodivergent users for whom sensory and communicative norms differ significantly, and LGBTQ+ users whose relational scripts fall outside heteronormative frameworks are all rendered, in Fricker's terms, hermeneutically dispossessed [6]: the current fembot technology offers no interpretive resources through which their intimate lives can be recognised or served.

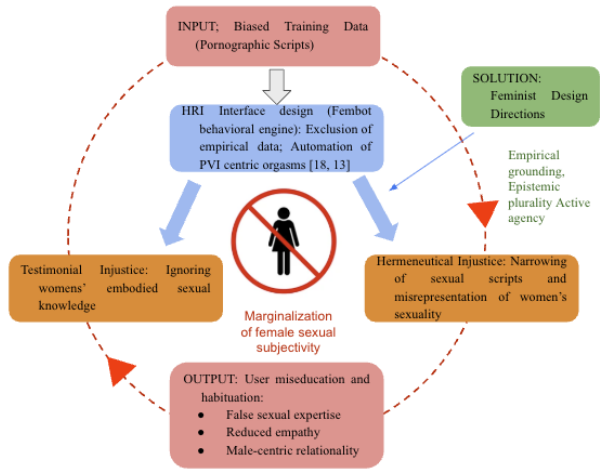

**Figure 1: The Epistemic Feedback Loop in Intimate HRI of Fembots**

## 4. FEMINIST DESIGN DIRECTIONS: TOWARD EPISTEMIC EQUITY IN INTIMATE HRI

For HRI to promote wellbeing, design must shift from "Submission" to "Situated Knowledge" [8]. This approach, grounded in Donna Haraway's philosophy, argues that all knowledge is produced from a specific social position rather than a view from nowhere — which provides the epistemological foundation for design that centers the lived experiences of those the current system excludes rather than treating male hedonistic preference as a universal standard.

## 4.1 Empirical Grounding: Prioritizing Scientific Truth over Pornographic Scripts

Current fembot interfaces are often built on sexual scripts designed to confirm male sexual success through exaggerated, porn-inspired simulations. To achieve epistemic justice, systems must be reprogrammed using data from empirical sexual health research rather than pornographic scripts [18, 13, 11].

**Physiological Accuracy:** Rather than programming 'orgasm on command' triggered solely by penetration, behavioural engines should be redesigned to reflect the evidence base on female sexual physiology [18, 13, 11]. Redesigned systems that accurately represent the physiological pathways to female pleasure can function as corrective tools for sexual ignorance rather than vehicles for its exacerbation.

**Evidence-Based Interaction:** Design must move away from hypersexualized performances toward a nuanced comprehension of sexual experiences that includes communication, consent, and individual preferences.

**Evaluation Criteria:** Empirical grounding can be assessed through content audits of behavioural scripts, measuring the proportion of programmed sexual responses derived from peer-reviewed sexual health research versus pornographic sources. User studies could additionally measure whether interaction with the system increases participants' accurate knowledge of female sexual physiology — for example, pre- and post-interaction assessments of understanding of female sexual anatomy and physiological pathways to orgasm — compared to a control condition.

## 4.2 Epistemic Plurality: Moving Beyond Male-centric Standards

Epistemic plurality demands that HRI design move away from hypersexualized, Western, penetration-normative and male-centric standards that exclude a significant portion of the human population [25, 23]. Following the work of Harding and Longino, robots should be reimagined as sites of curiosity and mutuality [28, 19, 20].

**Diverse Representation:** Fembots should reflect diverse bodies and sexualities, providing authentic representations of female sexuality and varied body types.

**Inclusive Interfaces:** By deconstructing traditional scripts, developers can create interfaces that recognize the unique needs and dignities of disabled, neurodivergent, and LGBTQ+ users [25].

**Gender Fluid Design:** Manufacturers should explore design alternatives that are not hyper-gendered, inviting "play" and "exploration" rather than dominance and control.

**Situated Knowledge:** Incorporating intersectional design principles — grounded in Haraway's Situated Knowledge framework [8] — ensures that technologies are shaped by those whose lives and sexualities have historically been erased or distorted in traditional knowledge production.

**Evaluation criteria:** Epistemic plurality can be operationalised through demographic representation audits of available body configurations, personality profiles, and interaction scripts — measuring diversity across race, body type, ability, and sexual orientation. Participatory design processes should be evaluated by tracking the proportion of co-designers drawn from marginalized communities, and post-deployment user satisfaction surveys should be disaggregated by gender identity, sexuality, and disability status to identify whether inclusive design goals are being met in practice.

## 4.3 Active Consent and Agency: Modeling Ethical Relationality

The "roboticization of consent" [9, 14] is a primary harm in current fembot design, where machines are programmed to be "always available" or to simulate "resistance-to-compliance" for indulging users' rape fantasies [14].

**Relational Agency:** Fembot interaction should not be designed as a one-way transaction. For fembots to be educational tools for human intimacy, developers must build systems in which engagement is conditional — the robot responds to communication, not simply command — modelling the mutual attentiveness that genuine intimacy requires [23].

**Habituation of Consent:** Instead of habituating users to a model of sex that ignores the reality of consent, robots should be programmed to require active communication and consent [26].

**Ethical Constraints:** Robots should be programmed with non-negotiable refusal states, interactions that the robot will not perform regardless of user instruction, such as simulating non-consensual scenarios or overriding a simulated "no." These hard constraints would structurally prevent the habituation of coercive sexual dynamics [27].

**Evaluation criteria:** Consent modelling can be assessed through behavioural audits that test whether override mechanisms for simulated refusal exist and can be triggered — their presence constituting a direct design failure.

## 5. IMPLEMENTATION CONSTRAINTS AND RESEARCH DIRECTIONS

The three feminist design directions proposed in Section 4 represent necessary reorientations of fembot HRI; however, translating them into implementable systems requires confronting significant real-world constraints across data collection, privacy, and governance.

**Responsible Data Collection:** Empirically grounding fembot behavioural engines demands data derived from peer-reviewed sexual health research rather than pornographic scripts [18, 13, 11]. However, collecting diverse, representative data on sexual experience raises serious ethical challenges. Participatory data collection with marginalised communities — disabled, LGBTQ+, and neurodivergent users — requires sustained, community-led consent processes, institutional ethics oversight, and clear protocols for data withdrawal [25, 28].

**Privacy and Safety Trade-offs:** Consent-modelling systems that prevent override of simulated refusal — proposed in Section 4.3 — may conflict with commercial pressures and existing regulatory vacuums in sextech [5, 27]. Implementing non-negotiable refusal states requires both technical enforcement and jurisdictional policy support [9, 27]. Moreover, designing for inclusivity across neurodivergent and disabled users may require adaptive interfaces that, without careful governance, could inadvertently expose vulnerability profiles [26]. These tradeoffs underscore that epistemic equity cannot be achieved through design alone: it requires privacy-by-design standards and community accountability mechanisms as implementation requirements [25, 28].

## 6. CONCLUSION

The current state of fembot AI is not a neutral reflection of desire but a technical reinforcement of epistemic injustice. This paper has addressed that problem across three levels. First, it applied Fricker's framework of testimonial and hermeneutical injustice [6] to fembot interface design, demonstrating that the harms encoded in these systems are not incidental design failures but structural epistemic ones — the systematic discrediting of women's embodied sexual knowledge [18, 13, 11] and the narrowing of the interpretive resources through which mutuality and consent can be understood [9, 14]. Second, it proposed three feminist design directions — empirical grounding, epistemic plurality, and active consent modelling — each mapped onto an identified harm and accompanied by concrete evaluation criteria that future researchers and developers can operationalise. Third, it stress-tested those directions against real-world implementation constraints, establishing that epistemic equity in intimate HRI requires not only redesigned fembots but privacy-by-design standards and community-led data governance [27, 26].

Equitable robotics for wellbeing demands that we reimagine the fembot not as a submissive object of male fantasy, but as a technology where the epistemic rights of all users are recognised and actively designed for [6, 8].

## REFERENCES

[1] J. Danaher. 2018. Should we be thinking about robot sex? In Robot Sex, J. Danaher and N. McArthur (Eds.). MIT Press, 3–14.

[2] C. Tonna-Barthet. 2018. The harmful effects of sex robots. Trinity Women & Gender Minorities Review 2, 1 (2018), 23–32.

[3] R. S. Eskens. 2017. Is sex with robots rape? Journal of Practical Ethics.

[4] K. R. Hanson and C. C. Locatelli. 2022. From sex robots and beyond: A narrative review. Current Sexual Health Reports 14, 3 (2022), 106–117.

[5] Grand View Research. 2024. SexTech Market Size, Share & Trends Analysis Report.

[6] M. Fricker. 2007. Epistemic Injustice: Power and the Ethics of Knowing. Oxford University Press.

[7] N. McArthur. 2018. The case for sexbots. In Robot Sex. MIT Press, 32–45.

[8] D. Haraway. 2013. Situated knowledges: The science question in feminism and the privilege of partial perspective. In M. Wyer, M. Barbercheck, D. Cookmeyer, H. Örün Öztürk, and M. Wayne (Eds.), Women, Science, and Technology: A Reader in Feminist Science Studies (pp. 455–472). Routledge.

[9] S. M. Gutiu. 2016. The roboticization of consent. In Robot Law, R. Calo et al. (Eds.). Edward Elgar Publishing, 186–212.

[10] D. Levy. 2007. Love and Sex with Robots: The Evolution of Human-Robot Relationships. Harper.

[11] E. A. Mahar, L. B. Mintz, and B. M. Akers. 2020. Orgasm equality: Scientific findings and societal implications. Current Sexual Health Reports 12, 1 (2020), 24–32.

[12] L. Williams. 2004. Porn Studies. Duke University Press.

[13] S. Hite. 2005. The Hite Report: A Nationwide Study of Female Sexuality (Reprint of 1976 ed.). Seven Stories Press, New York.

[14] E. M. Kaufman. 2020. Reprogramming consent: Implications of sexual relationships with artificially intelligent partners. Psychology & Sexuality 11, 4 (2020), 372–383.

[15] K. Richardson. 2016. Sex robot matters. IEEE Technology and Society Magazine 35, 2 (2016), 46–53.

[16] T. Oleksy and A. Wnuk. 2021. Do women perceive sex robots as threatening? Computers in Human Behavior 117.

[17] R. Sparrow. 2017. Robots, rape, and representation. (as cited in Eskens).

[18] E. A. Armstrong, P. England, and A. C. K. Fogarty. 2012. Accounting for women's orgasm and sexual enjoyment in college hookups and relationships. American Sociological Review 77, 3 (2012), 435–462.

[19] S. Harding. 1990. Whose Science? Whose Knowledge? Thinking from Women's Lives. Cornell University Press.

[20] H. E. Longino. 2017. Feminist epistemology. In The Blackwell Guide to Epistemology. Wiley, 325–353.

[21] N. Andrejek, T. Fetner, and M. Heath. 2022. Climax as work: Heteronormativity, gender labor, and the gap in orgasms. Gender & Society 36, 2 (2022), 189–213.

[22] G. M. Wetzel. 2023. Challenging biological justifications for the orgasm gap. Rutgers University.

[23] A. Srinivasan. 2021. The Right to Sex. Farrar, Straus and Giroux.

[24] B. Fahs. 2014. Coming to power: Women's fake orgasms and best orgasm experiences illuminate the failures of (hetero)sex. Culture, Health & Sexuality 16, 8 (2014), 974–988.

[25] R. Benjamin. 2019. Race After Technology: Abolitionist Tools for the New Jim Code. Polity, Cambridge, UK.

[26] A. Peeters and P. Haselager. 2021. Designing virtuous sex robots. International Journal of Social Robotics 13, 1 (2021), 55–66.

[27] L. Frank and S. Nyholm. 2017. Robot sex and consent. Artificial Intelligence and Law 25, 3 (2017), 305–323.

[28] J. Loh and M. Coeckelbergh. 2019. Feminist Philosophy of Technology. J.B. Metzler.